\documentclass[runningheads,a4paper]{llncs}
\usepackage{amssymb}
\usepackage{graphicx}
\usepackage{url}
\usepackage{hyperref}
\urldef{\mails}\path|{bxu, rdefrein, erobson, mofoghlu}@tssg.org| 
\newcommand{\keywords}[1]{\par\addvspace\baselineskip
\noindent\keywordname\enspace\ignorespaces#1}
 
\usepackage{amsmath}
\usepackage{algorithm}
\usepackage{algorithmic}
\usepackage{multirow}

\newtheorem{Property1}{Property}
\newtheorem{TheoremGanter}{Theorem}
\newtheorem{Property5}[Property1]{Property}
\newtheorem{Theorem6}[TheoremGanter]{Theorem}
\newtheorem{Theorem7}[TheoremGanter]{Theorem}
\usepackage{float,lipsum}
\floatstyle{boxed}
\newfloat{boxx}{thb}{lob}
\floatname{boxx}{Example}

\newtheorem{rdf8}{Example}

\makeatletter 
  \newcommand\figcaption{\def\@captype{figure}\caption} 
  \newcommand\tabcaption{\def\@captype{table}\caption} 
\makeatother

\begin{document}
\mainmatter
\title{Distributed Formal Concept Analysis Algorithms Based on an Iterative MapReduce Framework}
\titlerunning{Distributed FCA Algorithms: MRGanter}
\author{Biao Xu\and Ruair\'{i} de Fr\'{e}in\and Eric Robson\and M\'{i}che\'{a}l \'{O} Foghl\'{u}}
\authorrunning{Xu. et al,}
\institute{Telecommunications Software \& Systems Group,\\
Waterford Institute of Technology, Ireland\\
\mails\\}
\maketitle

\begin{abstract}
While many existing formal concept analysis algorithms are efficient, they are typically unsuitable for distributed implementation. Taking the MapReduce (MR) framework as our inspiration we introduce a distributed approach for performing formal concept mining. Our method has its novelty in that we use a light-weight MapReduce runtime called Twister which is better suited to iterative algorithms than recent distributed approaches. First, we describe the theoretical foundations underpinning our distributed formal concept analysis approach. Second, we provide a representative exemplar of how a classic centralized algorithm can be implemented in a distributed fashion using our methodology: we modify Ganter's classic algorithm by introducing a family of $\mbox{MR}^\star$ algorithms, namely MRGanter and MRGanter+ where the prefix denotes the algorithm's lineage. To evaluate the factors that impact distributed algorithm performance, we compare our $\mbox{MR}^{*}$ algorithms with the state-of-the-art. Experiments conducted on real datasets demonstrate that MRGanter+ is efficient, scalable and an appealing algorithm for distributed problems.
\end{abstract}

\section{Introduction}
Formal Concept Analysis (FCA), pioneered in the 80's by Wille~\cite{wille82restructuring}, is a method for extracting formal concepts --natural clusters of objects and attributes-- from binary object-attribute relational data. FCA has great appeal in the context of knowledge discovery \cite{Lakhal05ARMFCA}, information retrieval~\cite{fca2007inforetr} and social networking analysis applications \cite{snafca2009} because arranging data as a concept lattice yields a powerful and intuitive representation of the dataset \cite{wille82restructuring,lattice2003caspard}. 

FCA relies on closure operation which searches implication of attributes (obejcts) \cite{FCAMF1999Ganter}. According to this property, new formal concepts may be extracted iteratively by mapping a set of attributes (objects). While existing FCA algorithms perform this procedure iteratively and needs to access datasets each iteration. They are appropriate to process small centralized datasets. The recent explosion in dataset sizes, privacy protection concerns, and the distributed nature of the systems that collect this data, suggests that efficient distributed FCA algorithms are required. In this paper we introduce a distributed FCA approach based on a light-weight MapReduce runtime called Twister~\cite{twister10}, which is suited to iterative algorithms, scales well and reduces communication overhead.

\subsection{Related Work}
Some well-known algorithms for performing FCA include Ganter's algorithm~\cite{ganter1984two}, Lindig's algorithm~\cite{lindig2000fast} and CloseByOne~\cite{ClosebyOne1993Kuznetsov,InClose2009Andrews} and their variants~\cite{vychodil2008new,PCbO2008}. Ganter introduces \emph{lectic} ordering so that not all potential attribute subsets of the data have to be scanned when performing FCA. Ganter's algorithm computes concepts iteratively based on the previous concept without incurring exponential memory requirements.
In contrast, CloseByOne produces many concepts in each iteration. Bordat's algorithm~\cite{bordat1986} runs in almost the same amount of time as Ganter's algorithm, however, it takes a local concept generation approach. Bordat's algorithm introduces a data structure to store previously found concepts, which results in considerable time savings. Berry proposes an efficient algorithm based on Bordat's approach which require a data structure of exponential size~\cite{berry2006local}. A comparison of theoretical and empirical complexity of many well-known FCA algorithms is given in \cite{kuznetsov02comparing}. In addition, some useful principles for evaluating algorithm performance for sparse and dense data are suggested by Kuznetsov and Obiedkov; We consider data density when evaluating our approach.

The main disadvantage of the batch algorithms discussed above is that they require that the entire lattice is reconstructed from scratch if the database changes. Incremental algorithms address this problem by updating the lattice structure when a new object is added to database. Incremental approaches have been made popular by Norris~\cite{Norris1978}, Dowling~\cite{Dowling1993}, Godin et al.~\cite{godin1995incremental}, Capineto and Romano~\cite{carpineto1996}, Valtchev et al.~\cite{valtchev2002incremental} and Yu et al.~\cite{ImAddInc09}. In recent years, to reduce concept enumeration time, some parallel and distributed algorithms have been proposed. Krajca et al. proposed a parallel version based on CloseByOne~\cite{PCbO2008}. The first distributed  algorithm~\cite{DFCA2009MAPR} was developed by Krajca and Vychodil in 2009 using the MapReduce framework~\cite{MapReduce_Google}. In order to encourage more wide-spread usage of FCA, beyond the traditional FCA audience, we propose the development and implementation of efficient, distributed FCA algorithms. Distributed FCA is particularly appealing as distributed approaches that can potentially take advantage of cloud infrastructures to reduce enumeration time perhaps, are attractive to practitioners.
 
\subsection{Contributions}
We utilize the MapReduce framework in this paper to execute distributed algorithms on different nodes. Several implementations of MapReduce have been developed by a number of companies and organizations, such as Hadoop MapReduce by Apache\footnote{http://hadoop.apache.org/mapreduce/}, 
and Twister Iterative MapReduce\footnote{http://www.iterativemapreduce.org/}, since its inception by Google in 2004. A crucial distinction between the present paper and the work of Krajca and Vychodil~\cite{DFCA2009MAPR} is that we use a Twister implementation of MapReduce. 
Twister supports iterative algorithms~\cite{twister10}: we leverage this property to reduce the computation time of our  distributed FCA algorithms. In contrast, Hadoop architecture is designed for performing single step MapReduce.  We implement new distributed versions (MRGanter and MRGanter+) of Ganter's algorithm and empirically evaluate their performance. In order to provide an established and credible benchmark under equivalent experimental conditions, MRCbo, the distributed version of CloseByOne is implemented as well using Twister.

This paper is organized as follows. Section~\ref{sec:fca} gives an overview of Formal Concept Analysis and Ganter's algorithm. The  theoretical underpinnings for implementing FCA using distributed datasets are described in  Section~\ref{section:distriAlg} to support our approach. Our main contribution is a set of Twister-based distributed versions of Ganter's algorithm. Section~\ref{sec:twister} presents an implementation overview and comparison of Twister and Hadoop MapReduce. Empirical evaluation of the algorithms proposed in this paper is performed using real datasets from the UCI KDD machine learning repository, and experimental results are discussed in Section~\ref{section:experiments}. In summary, MRGanter+ performs favourably in comparison to centralized versions.   

\section{Formal Concept Analysis}
\label{sec:fca}
We continue by introducing the notational conventions used in the sequel. Let $O$ and $P$ denote a finite set of objects and attributes respectively. The data ensemble, $S$, may be arranged in Boolean matrix form as follows: the objects and attributes are listed along the rows and columns of the matrix respectively; The symbol $\times$ is entered in a row-column position to denote 
an object has that attribute; An empty entry denotes that the object does not have that attribute. Formally, this matrix describes the binary relation between the sets $O$ and $P$. The object set $X$ has attribute set $Y$ if $(X, Y) \in I$, $X\in O$ and $Y\in P$. The triple $(O,P,I)$ is called a formal context.
For example, in Table~\ref{table:example},  $O=\{1,2,3,4,5,6\}$ and $P=\{a,b,c,d,e,f,g\}$, thus object $\{2\}$ has attributes $\{a,c,e,g\}$.
\begin{table}[t]
\caption{The symbol $\times$ indicates that an object has the corresponding attribute.}
\label{table:example}
\begin{center}
\begin{tabular}{c|c|c|c|c|c|c|c|}
    & ~a~ & ~b~ & ~c~ & ~d~ & ~e~ & ~f~  & ~g~ \\   
\hline  ~1~& $\times$ & $\times$ &   & $\times$ &   & $\times$ &   \\ 
\hline  ~2~& $\times$ &   & $\times$ &   & $\times$ &   & $\times$  \\ 
\hline  ~3~&   & $\times$ & $\times$  & $\times$ &   & $\times$  & $\times$ \\ 
\hline  ~4~&   & $\times$ &   & $\times$ & $\times$ &   &   \\ 
\hline  ~5~& $\times$ &   &   & $\times$ & $\times$ & $\times$ &   \\ 
\hline  ~6~&   & $\times$ & $\times$ &   &   & $\times$ & $\times$ \\ 
\hline 
\end{tabular}
\end{center}
\end{table} 

We define a derivation operator on $X$ and $Y$ where $X\subseteq O$ and $Y\subseteq P$ as:
\begin{eqnarray}
\label{equ21}
X^{\prime} &=& \{p \in P \mid \, \forall t\in O:(t,p) \in I\} \\
Y^{\prime} &=& \{t \in O \mid \, \forall p \in P:(t,p) \in I\}.
\label{equ22}
\end{eqnarray}
The operation $X^{\prime}$ generates the set of attributes which are common to all objects in $X$. Similarly, $Y^{\prime}$ generates the set of all objects which are common to all attributes in $Y$. A pair $\langle X,Y\rangle$ is called a formal concept of $(O,P,I)$ if and only if $X\subseteq O$, $Y\subseteq P$, $X^{\prime}=Y$, and $Y^{\prime}=X$, where $X$ and $Y$ are called its extent and intent. The crucial property of a formal concept is that the mappings $X \mapsto X^{\prime\prime}$ and $Y \mapsto Y^{\prime \prime}$, commonly known as \emph{closure operators}, hold. The closure operator is used to calculate the extent and intent that form a formal concept.

In the following sections we describe established algorithms for concept mining, namely Ganter's algorithm (also known as NextClosure) and CloseByOne. We then introduce our distributed extensions of these approaches. 

\subsection{Ganter: Iterative Closure Mining Algorithm}
\label{subsec:ganter}
The NextClosure algorithm describes a method for generating new closures which guarantees every closure is enumerated once. Closures are generated iteratively using a pre-defined order, namely lectic ordering. The set of all formal concepts is denoted by $\mathcal{F}$. Let us arrange the elements of $P = \{p_{1}, \cdots, p_{i}, \cdots, p_{m}\}$ in an arbitrary linear order $p_{1}<p_{2}<\cdots <p_{i}<\ldots <p_{m}$, where $m$ is the cardinality of the attribute  set, $P$. The decision to use lectic ordering dictates that any arbitrarily chosen subset of $P$ is also ordered according to the \textit{lectic} ordering which was defined \emph{ab initio}. Given two subsets $Y_{1}$, $Y_{2}\subseteq P$, $Y_{1}$ is lectically smaller than $Y_{2}$ if the smallest element in which $Y_{1}$ and $Y_{2}$ differ belongs to $Y_{2}$.
\begin{equation}
\label{equ:lectic1}
Y_{1}\leq Y_{2} : \Longleftrightarrow
 \exists _{p_i} (p_i\in Y_{2}, p_i\notin Y_{1}, \forall _{p_j<p_i}(p_j\in Y_{1} \Longleftrightarrow p_j\in Y_{2})). 
\end{equation}
NextClosure uses Eqn.~(\ref{equ:lectic1}) as a feasibility condition for accepting new candidate  formal concepts. Typically this difference in set membership is made more explicit by denoting the smallest element, $p_i,$ in which the set $Y_{1}$ and $Y_{2}$ differ.
\begin{equation}
\label{equ:lecticformal}
Y_{1}\leq _{p_i} Y_{2} : \Longleftrightarrow 
 \exists _{p_i} (p_i\in Y_{2}, p_i\notin Y_{1}, \forall _{p_j<p_i}(p_j\in Y_{1} \Longleftrightarrow p_j\in Y_{2})).
\end{equation}
To fix ideas, if the order of $P=\{a,b,c,d,e,f,g\}$ is defined as $a<b<c<d<e<f<g$, and two subsets of $P$, or \emph{itemsets}, $Y_{1}=\{a,c,e,g\}$ and $Y_{2}=\{a,b,e,g\}$ are examined then $Y_{1}\leq Y_{2}$ because the smallest element in which the two sets differ is $b$ and this element belongs to $Y_{2}$.

In general, each subset $Y\subseteq P$ may yield a closure, $Y^{\prime \prime}\subseteq P$; The NextClosure algorithm attempts to find all closures systematically by exploiting lectic ordering. The generative operation is the $\oplus$-operation: a new intent is generated by applying $\oplus$ on an existing intent and an attribute. 
Let the ordering of $P$ be $p_{1}<p_{2}<\ldots <p_{i}<\ldots <p_{m}$, and consider the subset $Y\subseteq P$. The $\oplus$-operator is defined as:
\begin{equation}
\label{ganter:oplus}
Y \oplus p_i := ((Y \cap \{p_1, \ldots, p_{i-1}\}) \cup \{p_i\})^ {\prime\prime}, \quad
\mbox{where }Y \subseteq P\mbox{ and }p_i\subset P.
\end{equation}
NextClosure then compares the new candidate formal concept with the previous concept. If the condition in Eqn.~(\ref{equ:lecticformal}) is satisfied the candidate concept produced by Eqn.~(\ref{ganter:oplus}) is kept.

The $\oplus$-operator in Eqn.~(\ref{ganter:oplus}) consists of intersection, union and closure operations; Lectic ordering and the associated complexity of these operations explains why NextClosure's ordered approach incurs high computational expense, and consequently why the largest dataset-size NextClosure can practically process is relatively small. 
\begin{rdf8}
\label{example_compute_all_concepts}
Consider the formal context in Table~\ref{table:example}. Assume we have a concept $\langle\{1,5\},\{a,d,f\}\rangle$. We take the attribute set, $Y=\{a,d,f\}$, and calculate, $Y\oplus e$. First, we compute, $\{a,d,f\}\cap \{a,b,c,d\}=\{a,d\}$, then we append $e$ and generate $\{a,d\}\cup \{e\}=\{a,d,e\}$. Performing $\{a,d,f\}\oplus e = \{a,d,e\}^{\prime \prime}$ yields the set, $\{a,d,e,f\}$. To demonstrate the role of lectic ordering, we compute  $Y\oplus c = \{a,c,e\}$. According to the feasibility condition in (Eqn.~\ref{equ:lecticformal}), $\{a,d,e,f\}\leq _{c}\{a,c,e\}$. Thus, the set, $\{a,c,e\}$, is added to the concept lattice, $\mathcal{F}$. By repeating this process, NextClosure determines that there are $21$ formal concepts in the concept lattice representation of the formal context in Table~\ref{table:example}. The set of concepts, $\mathcal{F}$, is listed in Table~\ref{example:concepts}.
\begin{table}[t]
\centering
\caption{Formal concepts mined from Table~\ref{table:example}, including empty concepts.}
\label{example:concepts}
\begin{tabular}{llllll}
$F_{1}$: & $\langle\{1,2,3,4,5,6\},\{\}\rangle$ & $F_{8}$: & $\langle\{1,3,4,6\},\{b\}\rangle$ & $F_{15}$: & $\langle\{1,2,5\},\{a\}\rangle$ \\ 
$F_{2}$: & $\langle\{1,3,5,6\},\{f\}\rangle$ & $F_{9}$: & $\langle\{1,3,6\},\{b,f\}\rangle$ & $F_{16}$: & $\langle\{2,5\},\{a,e\}\rangle$ \\ 
$F_{3}$: & $\langle\{2,4,5\},\{e\}\rangle$ & $F_{10}$: & $\langle\{1,3,4\},\{b,d\}\rangle$ & $F_{17}$: & $\langle\{1,5\},\{a,d,f\}\rangle$ \\ 
$F_{4}$: & $\langle\{1,3,4,5\},\{d\}\rangle$ & $F_{11}$: & $\langle\{1,3\},\{b,d,f\}\rangle$ & $F_{18}$: & $\langle\{5\},\{a,d,e,f\}\rangle$ \\ 
$F_{5}$: & $\langle\{1,3,5\},\{d,f\}\rangle$ & $F_{12}$: & $\langle\{4\},\{b,d,e\}\rangle$ & $F_{19}$: & $\langle\{2\},\{a,c,e,g\}\rangle$ \\ 
$F_{6}$: & $\langle\{4,5\},\{d,e\}\rangle$ & $F_{13}$: & $\langle\{3~ 6\},\{b,c,f,g\}\rangle$ & $F_{20}$: & $\langle\{1\},\{a,b,d,f\}\rangle$ \\ 
$F_{7}$: & $\langle\{2,3,6\},\{c,g\}\rangle$ & $F_{14}$: & $\langle\{3\},\{b,c,d,f,g\}\rangle$ & $F_{21}$: & $\langle\{\},\{a,b,c,d,e,f,g\}\rangle$ \\ 
\end{tabular}
\end{table}
\end{rdf8}

Pseudo code for NextClosure is described in the Algorithm~\ref{AC} and \ref{NC} as background to our distributed approach. Algorithm \ref{AC} applies the closure operator on the null attribute set and generates the first intent, $Y$, which is the base for all subsequent formal concepts. New concepts are generated in turn by calling Algorithm~\ref{NC} and concatenating the resultant concepts to the set of formal concepts, $\mathcal{F}$. As each candidate intent is extended with new attributes, the last intent should be the complete set of attributes. This feature is used to terminate the loop (in Line \ref{code_ganter_while} of the Algorithm~\ref{AC}). Algorithm \ref{NC} accepts the formal context triple, $(O,P,I)$ and current intent, $Y$, as inputs. By convention, the attribute set $P$ is sorted in descending order. The $\oplus$-operator described in Eqn.~\ref{ganter:oplus} is applied to produce candidate formal concepts. The concept feasibility condition Eqn.~(\ref{equ:lecticformal}) is used to verify whether a new candidate should be added to the set of formal concepts, $\mathcal{F}$.
\begin{figure*}[t]
\begin{minipage}[t]{.45\textwidth}
\begin{algorithm}[H]
\caption{AllClosure} \label{AC}
\begin{algorithmic}[1]
\scriptsize
\REQUIRE $\emptyset$: null attribute set. 
\ENSURE $\mathcal{F}$: Formal concepts set.
\STATE $Y \leftarrow \emptyset ^{\prime \prime}$;
\WHILE{$Y$ is not the last closure} \label{code_ganter_while}
	\STATE $Y$ $\leftarrow$ NextClosure();
	\STATE $\mathcal{F} \leftarrow \mathcal{F} \cup Y$;
\ENDWHILE \label{code_ganter_while_end}
\STATE \textbf{return} $\mathcal{F}$
\end{algorithmic}
\end{algorithm}
\end{minipage}
\hspace{2mm}
\begin{minipage}[t]{.45\textwidth}
\begin{algorithm}[H]
\caption{NextClosure} \label{NC}
\begin{algorithmic}[1]
\scriptsize
\REQUIRE 
$O,P,I,Y$: formal context \& current intent.
\ENSURE $Y$.
\FOR{$p_{i}$ from $p_{m}$ down to $p_{1}$} \label{code_ganter_for}
	\IF{$p_i \notin Y$}
		\STATE \mbox{candidate} $\leftarrow Y \oplus p_{i}$; \label{code_ganter_OPlus}
		\IF{\mbox{candidate} $\leq_{p_i} Y$} \label{code_ganter_condition}
			\STATE $Y \leftarrow$ \mbox{candidate};
			\STATE break;
		\ENDIF
	\ENDIF
\ENDFOR
\STATE \textbf{return} $Y$
\end{algorithmic}
\end{algorithm}
\end{minipage}
\end{figure*}
The approach taken in the CloseByOne algorithm is similar in spirit to the approach taken by the NextClosure algorithm: CloseByOne generates new formal concepts based on concept(s) generated in the previous iteration and tests their feasibility using the operator, $\leq_{p_{i}}$. The crucial difference is that the CloseByOne algorithm generates many concepts in each iteration. CloseByOne terminates when there are no more concepts that satisfy Eqn.~(\ref{equ:lecticformal}). In short, NextClosure only finds the first feasible formal concept in each iteration whereas CloseByOne potentially generates many. As a consequence, CloseByOne requires far fewer iterations.

The appeal of NextClosure, and explanation for our desire to make it more efficient lies in its thoroughness; the guarantee of a complete lattice structure which is a consequence of the main theorem of Formal Concept Analysis~\cite{FCAMF1999Ganter}. This thoroughness is due to lectic ordering and the iterative approach deployed by NextClosure; however, thoroughness comes at the cost of high complexity. 
The advent of efficient mechanisms for dealing with iterative algorithms using MapReduce  captured by Twister allow us to couple NextClosure's thoroughness with a practical distributed implementation in this paper. 

\section{Distributed Algorithms for Formal Concept Mining}
\label{section:distriAlg}
We continue by describing two methods for performing distributed NextClosure, namely, MRGanter and MRGanter+. An introduction to Twister is deferred to Section~\ref{sec:twister}. We start by describing the properties of a partitioned dataset compared to its unpartitioned form. In many cases these properties are simply restatements of the properties of the derivations operators.

Given a dataset $S$, we partition its objects into $n$ subsets and distribute the subsets over $n$ different nodes. Without loss of generality, it is convenient to limit $n=2$ here. We denote the partitions by $S_{1}$ and $S_{2}$. Alternatively we can think in terms of formal contexts and write the formal context, $(O,P,I)$, in terms of the partitioned formal contexts $(O_{S_{1}},P,I_{S_{1}})$ and $(O_{S_{2}},P,I_{S_{2}})$. To fix ideas, we use the dataset in Table~\ref{table:example} as an exemplar and generate the partitions in Table~\ref{distributeddataset}. The partitions are non-overlapping: the intersection of the partitions is the null set, $S_{1}\cap S_{2}=\emptyset$ and their union gives the full dataset $S = S_{1}\cup S_{2}$. It follows that the partitions, $S_{1}$, $S_{2}$, have the same attributes sets, $P$, as the entire dataset $S$, however, the set of objects is different in each partition, e.g. $O_{S_{1}}$ and $O_{S_{2}}$. 
\begin{table}[t]
\centering
\caption{Partitioned datasets $S_{1}$ and $S_{2}$ derived from Table~\ref{table:example}}
\begin{tabular}{lr}
\begin{tabular}{|lc|c|c|c|c|c|c|} 
\multicolumn{8}{c}{$S_{1}$ or $(O_{S_{1}},P,I_{S_{1}})$} \\ \hline
     &~a~ & ~b~ & ~c~ & ~d~ & ~e~ & ~f~  & ~g~ \\   
\hline  ~1~\vline & $\times$ & $\times$ &   & $\times$ &   & $\times$ &   \\ 
\hline  ~2~\vline & $\times$ &   & $\times$ &   & $\times$ &   & $\times$  \\ 
\hline  ~3~\vline &   & $\times$ & $\times$  & $\times$ &   & $\times$  & $\times$ \\ 
\hline
\end{tabular} & 
\hspace{4mm} 
\begin{tabular}{|lc|c|c|c|c|c|c|}
\multicolumn{8}{c}{$S_{2}$ or $(O_{S_{2}},P,I_{S_{2}})$} \\ \hline
 & ~a~ & ~b~ & ~c~ & ~d~ & ~e~ & ~f~  & ~g~ \\  
\hline  ~4~\vline &   & $\times$ &   & $\times$ & $\times$ &   &   \\ 
\hline  ~5~\vline & $\times$ &   &   & $\times$ & $\times$ & $\times$ &   \\ 
\hline  ~6~\vline &   & $\times$ & $\times$ &   &   & $\times$ & $\times$ \\ 
\hline 
\end{tabular}
\end{tabular}
\label{distributeddataset}
\end{table} 

Let $Y_{S}$, $Y_{S_1}$ and $Y_{S_2}$ denote an arbitrary attribute set $Y$ with respect to the entire dataset $S$, and each of its partitions $S_1$ and $S_2$ respectively. By construction they are equivalent: $Y_{S} \equiv Y_{S_1} \equiv Y_{S_2}$. Similarly, $Y'_{S}$, $Y'_{S_1}$ and $Y'_{S_2}$ are the sets of objects derived by the derivation operation in each of the partitions $S_1$, $S_2$ and the entire dataset $S$ respectively. 
\begin{Property1}
\label{lemma1}
Given the formal context, $(O,P,I)$, the two partitions $(O_{S_{1}},P,I_{S_{1}})$ and $(O_{S_{2}},P,I_{S_{2}})$, we have the property $Y_S^{\prime} = Y_{S_1}^{\prime} \cup Y_{S_2}^{\prime}:$ the union of the sets of objects generated by the derivation of the attribute sets $Y_{S_1}$ and $Y_{S_2}$ overs the partitions is equivalent to the set of objects generated by the derivation of the attribute set $Y_{S}$ over the entire dataset, $S$. 
\end{Property1}
Appealing to the definition of the derivation operator proposed by Wille in~\cite{wille82restructuring}, the set, $Y^{\prime}_{S}$, is a subset of $O$, $Y^{\prime}_{S} \subseteq O$. Moreover, $Y^{\prime}_{S_1} \subseteq O_{S_{1}}$ and $Y^{\prime}_{S_2} \subseteq O_{S_{2}}$. Given  $S_{1}\cup S_{2}=S$ and $S_{1}\cap S_{2}=\emptyset$, it follows that, $O_{S_{1}} \cup O_{S_{2}} = O$ and $O_{S_{1}} \cap O_{S_{2}} = \emptyset$; Therefore,  $Y_{S_{1}}^{\prime} \subseteq Y_{S}^{\prime}$ and $Y_{S_{2}}^{\prime} \subseteq Y_{S}^{\prime}$. Finally, $Y_{S_{1}}^{\prime} \cup Y_{S_{2}}^{\prime}\equiv Y_{S}^{\prime}$. 
As a counterexample, an object $t$ that exists in $Y_{S}^{\prime}$, but not in $Y_{S_{1}}^{\prime}$ or $Y_{S_{2}}^{\prime}$, cannot exist because $O_{S_{1}} \cup O_{S_{2}} = O$ and $O_{S_{1}} \cap O_{S_{2}} = \emptyset$ and $Y_{S}=Y_{S_{1}}=Y_{S_{2}}$. If $t$ is in $Y_S'$ it must appear in $Y_{S_{1}}^{\prime}$ or $Y_{S_{2}}^{\prime}$. In short, Property~\ref{lemma1} allows us to process all objects independently: the objects can be distributed and processed in an arbitrary order and this will not affect the result of $Y^{\prime}$. Property~\ref{lemma1} is trivially extended to the case of $n$ partitions. Now we describe how formal concepts can be combined from different partitions.
\begin{Property5}
\label{theorem5}
Given the formal context, $(O,P,I)$, the two partitions $(O_{S_{1}},P,I_{S_{1}})$ and $(O_{S_{2}},P,I_{S_{2}})$, we have the property  $Y_S^{\prime\prime} = Y^{\prime\prime}_{S_1} \cap Y^{\prime\prime}_{S_2}:$ The intersection of the closures of the attribute set, $Y_{S_1}$ and $Y_{S_2}$, with respect to the partitions $S_1$ and $S_2$ is equivalent to the closure of the  attribute set, $Y_S$, with respect to the entire dataset $S$.
\end{Property5}
By the definition of the partition construction method above, $S_1 \cup S_2 = S$, which implies that, $S_1 \subset S$ and $S_2 \subset S$. Recall that, $Y^{\prime}_{S_1} \subset Y^{\prime}_S$ and $Y^{\prime}_{S_2} \subset Y^{\prime}_S$, and from Property~\ref{lemma1} we have that $Y_S^{\prime} = Y_{S_1}^{\prime} \cup Y_{S_2}^{\prime}.$ Appealing to the properties of the derivation operators, in~\cite{wille82restructuring}, we have, $Y^{\prime\prime}_{S_1} \supseteq Y^{\prime\prime}_S$ and $Y^{\prime\prime}_{S_2} \supseteq Y^{\prime\prime}_S$. It is clear that $Y^{\prime\prime}_{S_{1}}$ and $Y^{\prime\prime}_{S_{2}}$ need not equal $Y^{\prime\prime}_S$, but by the definition of a closure $(Y^{\prime}_{S_1} \cup Y^{\prime}_{S_2})^{\prime} = (Y^{\prime}_S)^{\prime} = Y_S$, thus, $(Y^{\prime}_{S_1} \cup Y^{\prime}_{S_2})^{\prime} = 
Y^{\prime\prime}_{S_1} \cap Y^{\prime\prime}_{S_2}$ follows trivially from the definition of the derivations operators.
\begin{rdf8} \label{example:first} Consider the following example of Property~\ref{theorem5}.
Taking itemset $Y=\{b,d\}$. We derive $Y^{\prime\prime}_{S_1}=\{b,d,f\}$ from the first partition $S_1$, and $Y^{\prime\prime}_{S_2}=\{b,d,e\}$ from $S_2$. We derive $Y_S^{\prime\prime}=\{b,d\}$ for the entire dataset $S$. Therefore $Y_S^{\prime\prime}= Y^{\prime\prime}_{S_1} \cap Y^{\prime\prime}_{S_2}$.
\end{rdf8}
\begin{Theorem6}
\label{theorem6}
Given a set of attributes $Y$, $Y \subset P$. Let $\mathcal{F}^{Y}_{S_1}$ and $\mathcal{F}^{Y}_{S_2}$ be the sets of closures based on $Y$ in relation to $S_1$ and $S_2$ respectively. Then the closure set of $Y$ in relation to $S$ can be calculated from: 
$\mathcal{F}^{Y}_{S} = \mathcal{F}^{Y}_{S_1} \cap \mathcal{F}^{Y}_{S_2}$
\end{Theorem6}
This is simply a consequence of Property~\ref{theorem5} as, $\mathcal{F}^{Y}_{S} = Y^{\prime\prime}_{S} = Y^{\prime\prime}_{S_1} \cap Y^{\prime\prime}_{S_2} = \mathcal{F}^{Y}_{S_1} \cap \mathcal{F}^{Y}_{S_2}$
 and $Y_{S} \equiv Y_{S_1} \equiv Y_{S_2}$ by definition of the partition.
\begin{rdf8}
Consider again Example~\ref{example:first}. Appealing to Theorem~\ref{theorem6}, the formal concept with respect to the entire data set is the intersection of the formal concepts from each partition $F^{Y}_{S}=F^{Y}_{S_1} \cap F^{Y}_{S_2}=\mbox{\{b,d,f\}} \cap \mbox{\{b,d,e\}=\{b,d\}}$.
\end{rdf8}
We denote the $k$-th partition as $S_k$ where $k=1,\cdots,n$ and then propose:
\begin{Theorem7}
\label{theorem7}
Given the closures $\mathcal{F}^{Y}_{S_1}$, \ldots, $\mathcal{F}^{Y}_{S_n}$ from $n$ disjoint partitions, $\mathcal{F}^{Y}_{S} = \mathcal{F}^{Y}_{S_1} \cap \ldots \cap \mathcal{F}^{Y}_{S_n}$.
\end{Theorem7}
A trivial inductive argument establishes that Theorem~\ref{theorem7} holds. Theorem~\ref{theorem6} proves the $n = 2$ case. Theorem~\ref{theorem7} follows by recognizing that
the dataset $S$ at the $(k-1)$-th step of the proof can be thought as of consisting of  two partitions only, the partition ${S_1 \cup \cdots \cup S_{k-1}}$ and a second partition $S_k$.

Calling on nothing more complex than: 1) the properties of the derivation operators, and 2) construction of non-overlapping partitions, we leverage Theorem~\ref{theorem7} in order to apply the MapReduce, specifically the Twister variant, to calculate closures from arbitrary number of distributed nodes sure in the knowledge that the thoroughness of NextClosure is preserved.

\subsection{MRGanter}
\label{subsec:mrganter}
In order to address the dataset size limitations imposed on NextClosure --owing in particular to the complexity of the $\oplus$-operation-- we propose to deploy FCA across multiple nodes in order to reduce the computation time. We address the problem of how to decompose NextClosure so that each sub-task can be executed in parallel. In the Algorithm \ref{NC}, there were two stages involved in computing NextClosure: 1) computing a new candidate closure, and 2) making a judgement on whether to add it to the evaluated formal concepts. In MapReduce parlance, computing a new candidate closure corresponds to the \mbox{map} stage, and validating its feasibility corresponds to the \mbox{reduce} phase. For the purpose of this discussion, we only calculate the intent of a formal concept. In practice, we calculate an extent based on the intent and previous extent. The variables and constants used in these algorithms are summarized in Table~\ref{table:variables:DFCA}.
\begin{table}
\caption{Table of variables and constants for distributed FCA algorithms.}
\vspace{-5mm}
\label{table:variables:DFCA}
\begin{center}
\begin{tabular}{l|p{8.5cm}}
\hline 
Variables/Constants & Description \\ 
\hline 
\mbox{p\_i} & an attribute in \mbox{P}, where $i=1,\cdots,m$ \\ 
\mbox{L\_k} & the complete set of local closures in data partition $k$ where $k=1, \cdots, n$. It will be transfered from mapper to reducer \\ 
\mbox{l\_i} & an intent in \mbox{L\_k} which is derived from \mbox{p\_i} \\ 
\mbox{d} & the intent produced in the previous iteration \\
\mbox{f} & the newly generated intent  \\ 
\mbox{G} & a container for storing newly generated intents \\ \hline 
\end{tabular} 
\end{center}
\end{table}
\begin{figure*}[ttt!]
\begin{minipage}[t]{0.45\textwidth} 
\begin{algorithm}[H]
\caption{Merging function}
\label{merging}
\begin{algorithmic}[1]
\scriptsize
\REQUIRE \mbox{p\_i}, \mbox{L\_k}, \mbox{f}.\\
\ENSURE $f$.
\STATE \mbox{l\_i} $\leftarrow$ the local closure in \mbox{L\_k} in terms of $p\_i$;\label{merging_getAttribute}
\STATE \mbox{f} $\leftarrow \Psi$(\mbox{l\_i}, \mbox{f});
\STATE \textbf{return} \mbox{f}
\end{algorithmic}
\end{algorithm}
\end{minipage}
\hspace{4mm}
\begin{minipage}[t]{0.45\textwidth}
\begin{algorithm}[H]
\caption{Map: MRGanter}
\label{maptaskforMRGanter}
\begin{algorithmic}[1]
\scriptsize
\REQUIRE \mbox{d}.
\ENSURE \mbox{(d, L\_k)}.
\FOR{\mbox{p\_i} from \mbox{p\_m} down to \mbox{p\_1}}
	\IF{\mbox{p\_i} is not in \mbox{d}}
		\STATE \mbox{l\_i} $\leftarrow$ \mbox{d} $\oplus$ \mbox{p\_i};
		\STATE associate \mbox{l\_i} with \mbox{p\_i}; \label{mrganter:setAttribute}
		\STATE \mbox{L\_k} $\leftarrow$ \mbox{L\_k} $\cup$ \mbox{l\_i};
	\ENDIF
\RETURN \mbox{(d, L\_k)};
\ENDFOR
\end{algorithmic}
\end{algorithm}
\end{minipage}
\end{figure*}

The main operation in the merging function is the intersection operator, which is applied on the set of local closures \mbox{L\_k} generated at each node. Algorithm~\ref{merging} gives the pseudo code for the merging function based on Theorem~\ref{theorem7}. To describe the merging operation, we introduce the notation,  $\Psi($\mbox{l\_i, f}$)=$\mbox{l\_i} $\cap$ \mbox{f}, which acts on two intents. The merging function is deployed at the reduce phase and only processes the local closures derived from the same attribute (Line \ref{merging_getAttribute}).

The Map phase described in the Algorithm~\ref{maptaskforMRGanter} produces all local closures. The output consists of the previous intent \mbox{d} and a set of local intents \mbox{L\_k}. In order to be used in the merging function the attribute which was used to form local closures should be recorded and passed, as Line \ref{mrganter:setAttribute} does. All pairs which have the same key, \mbox{d}, will be sent to the same reducer. All local intents are used to form global intents in reduce phase.
\begin{figure*}[ttt!]
\begin{minipage}[t]{.45\textwidth}
\begin{algorithm}[H]
\caption{Reduce: MRGanter}
\label{reducetaskforMRGanter}
\begin{algorithmic}[1]
\scriptsize
\REQUIRE \mbox{(d,L\_k)}.\\
\ENSURE \mbox{f}.
\FOR{\mbox{p\_i} in \mbox{P}}
	\STATE \mbox{f} $\leftarrow$ initialize new intent;
	\FOR{$i$ from $1$ up to $m$}
		\STATE \mbox{f} $\leftarrow$ \mbox{merging(p\_i, L\_k, \mbox{f})}; \label{mrganter_merging}
	\ENDFOR
	\IF{\mbox{f} $\leq_{p\_i}$ \mbox{d}} \label{reduce_mrganter_condition}
		\STATE break;
	\ELSE
		\STATE \mbox{continue};
	\ENDIF 
\ENDFOR	
\RETURN \mbox{f}
\end{algorithmic}
\end{algorithm}
\end{minipage}
\hspace{4mm}
\begin{minipage}[t]{.45\textwidth}
\begin{algorithm}[H]
\caption{Reduce: MRGanter+}
\label{reducetaskforMRGanterPlus}
\begin{algorithmic}[1]
\scriptsize
\REQUIRE \mbox{(d, L\_k)}.
\ENSURE \mbox{G}.
\STATE \mbox{H} $\leftarrow$ initialize a two-level hash table;
\FOR{$p_i$ in $P$}
	\STATE \mbox{f} $\leftarrow$ initialize new intent;
	\FOR{$i$ from $1$ up to $m$}
		\STATE \mbox{f} $\leftarrow$ \mbox{merging(p\_i, L\_k, f)}; \label{mrganterplus_merging}
	\ENDFOR
	\IF{\mbox{f} is not in \mbox{H}} \label{mrganterplus_check}
		\STATE add \mbox{f} into \mbox{H};
	    \STATE add \mbox{f} into \mbox{G};
	\ENDIF
\ENDFOR \label{mrganterplus_for_end}
\STATE \textbf{return} \mbox{G}
\end{algorithmic}
\end{algorithm}
\end{minipage}
\end{figure*}

Algorithm \ref{reducetaskforMRGanter} accepts \mbox{(d,L\_k)} from the k-th \textit{mappers} (see Section~\ref{sec:twister}), where $k=1,\cdots,n$. Only pairs who have the same key, \mbox{d}, are accepted by a Reducer. Line \ref{mrganter_merging} generates an candidate closure \mbox{f}. This candidate is then validated. Finally, the successful candidate will be outputted as global closure \mbox{f}.

\begin{figure}[t]
\centering
\includegraphics[height=.38\textheight]{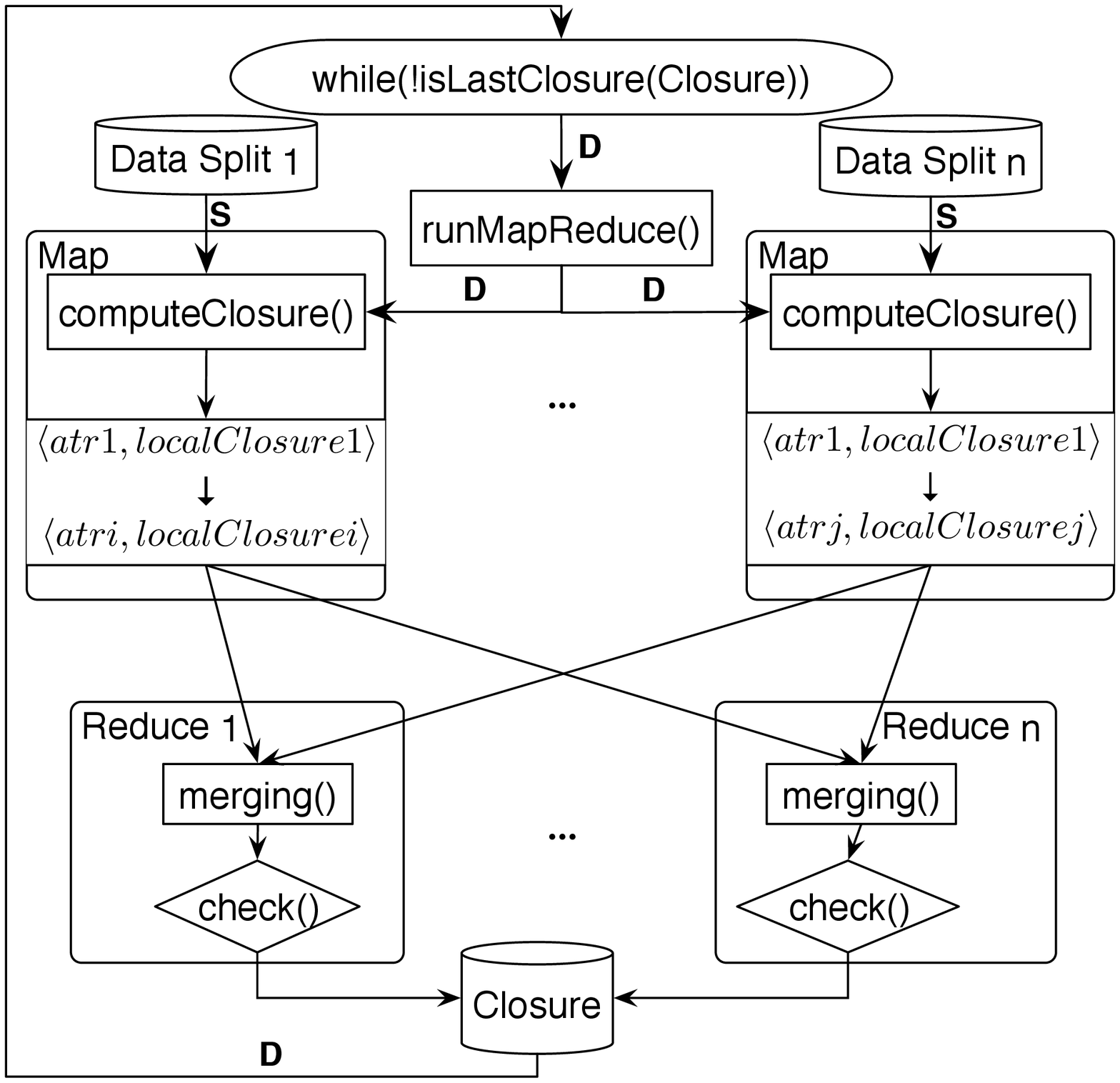}
\vspace{-3mm}
\caption{MRGanter work flow: static data is loaded at the start of the procedure (labeled by S) and the dynamic data (closures produced during each iteration) is passed and used in the next iteration (labeled by D).}
\label{fig:MRGanterIDF}
\end{figure}

Fig.~\ref{fig:MRGanterIDF} depicits the iterative flow of control of MRGanter; the lines marked with ``S'' import static data from each partition, while the lines marked with ``D'' configure each \mbox{map} with the previous closure. Each new  closure is tested to see if it is the last, e.g. it contains all attributes, $P$. If this condition is not met MRGanter continues. 

We present a worked example using the dataset in Table~\ref{distributeddataset}. Table~\ref{example_mrganter}, illustrates a few results due to space limitations. In practice, MRGanter performs 20 iterations to determine all concepts. 
\begin{table}[t]
\begin{minipage}[c]{0.46\textwidth}
\caption{MRGanter: In each iteration, only single a intent (bold) satisfies the condition.}
\label{example_mrganter}
\begin{tabular}{c|c|c|c|c}
\hline 
d & p\_i & l\_i from $S_{1}$ & l\_i from $S_{2}$ & f \\ 
\hline 
\multirow{4}{*}{$\emptyset$}
 & g & \{c,g\} & \{b,c,f,g\} & \{c,g\} \\ 
 & f & \{b,d,f\} & \{f\} & \textbf{\{f\}} \\ 
 & e & \{a,c,e,g\} & \{d,e\} & \{e\}\\ 
 & d & \{b,d,f\} & \{d,e\} & \{d\} \\
 & c & \{c,g\} & \{b,c,f,g\} & \{c,g\} \\ 
 & b & \{b,d,f\} & \{b\} & \{b\} \\
 & a & \{a\} & \{a,d,e,f\} & \{a\} \\
\hline 
\multirow{4}{*}{\{f\}}
 & g & \{b,c,d,f,g\} & \{b,c,f,g\} & \{b,c,f,g\} \\
 & e & \{a,c,e,g\} & \{d,e\} & \textbf{\{e\}} \\
 & d & \{b,d,f\} &  \{d,e\} & \{d\} \\
 & c & \{c,g\} &  \{b,c,f,g\} & \{c,g\} \\
 & b & \{b,d,f\} &  \{b\} & \{b\} \\
 & a & \{a\} & \{a,d,e,f\} & \{a\} \\
\hline
\multirow{4}{*}{\{e\}}
 & g & \{a,c,e,g\} & \{a,\ldots,g\} & \{a,c,e,g\} \\
 & f & \{a,\ldots,g\} &  \{a,d,e,f\} & \{a,d,e,f\} \\
 & d & \{b,d,f\} &  \{d,e\} & \textbf{\{d\}} \\
 & c & \{c,g\} &  \{b,c,f,g\} & \{c,g\} \\
 & b & \{b,d,f\} &  \{b\} & \{b\} \\
 & a & \{a\} & \{a,d,e,f\} & \{a\} \\
\hline
\multirow{4}{*}{\{d\}}
 & g & \{b,c,d,f,g\} &  \{a,\ldots,g\} & \{b,c,d,f,g\} \\
 & f & \{b,d,f\} & \{a,d,e,f\} & \textbf{\{d,f\}} \\
 & e & \{a,\ldots,g\} & \{d,e\} & \{d,e\} \\
 & c & \{c,g\} &  \{b,c,f,g\} & \{c,g\} \\
 & b & \{b,d,f\} & \{b\} & \{b\} \\
 & a & \{a\} & \{a,d,e,f\} & \{a\} \\
\hline
\end{tabular} 
\end{minipage}
\hspace{0.45cm}
\begin{minipage}[c]{0.46\textwidth}
\caption{MRGanter+: Many intents (bold) are maintained per iteration.}
\label{example_mrganter_plus}
\begin{tabular}{c|c|c|c|c}
\hline 
d & p\_i & l\_i from $S_{1}$ & l\_i from $S_{2}$ & f \\ 
\hline 
\multirow{4}{*}{$\emptyset$}
 & g & \{c,g\} & \{b,c,f,g\} & \textbf{\{c,g\}} \\ 
 & f & \{b,d,f\} & \{f\} & \textbf{\{f\}} \\ 
 & e & \{a,c,e,g\} & \{d,e\} & \textbf{\{e\}} \\ 
 & d & \{b,d,f\} & \{d,e & \textbf{\{d\}} \\ 
 & c & \{c,g\} & \{b,c,f,g\} & \{c,g\} \\ 
 & b & \{b,d,f\} & \{b\} & \textbf{\{b\}} \\
 & a & \{a\} & \{a,d,e,f\} & \textbf{\{a\}} \\
\hline
\multirow{4}{*}{\{cg\}}
 & f & \{b,c,d,f,g\} & \{b,c,f,g\} & \textbf{\{b,c,f,g\}} \\
 & e & \{a,c,e,g\} & \{a,\ldots,g\} & \textbf{\{a,c,e,g\}} \\
 & d & \{b,c,d,f,g\} & \{a,\ldots,g\} & \textbf{\{b,c,d,f,g\}} \\
 & b & \{b,d,f\} & \{b\} & \{b\} \\
 & a & \{a\} &  \{a,d,e,f\} & \{a\} \\
\hline 
\multirow{4}{*}{\{f\}}
 & g & \{b,c,d,f,g\} & \{b,c,f,g\} & \{b,c,f,g\} \\
 & e & \{a,c,e,g\} & \{d,e\} & \{e\} \\
 & d & \{b,d,f\} & \{d,e\} & \{d\} \\
 & c & \{c,g\} & \{b,c,f,g\} & \{c,g\} \\
 & b & \{b,d,f\} & \{b\} & \{b\} \\
 & a & \{a\} & \{a,d,e,f\} & \{a\} \\
\hline
\multirow{4}{*}{\{e\}}
 & g & \{a,c,e,g\} & \{a,\ldots,g\} & \{a,c,e,g\} \\
 & f & \{a,\ldots,g\} & \{a,d,e,f\} & \textbf{\{a,d,e,f\}} \\
 & d & \{b,d,f\} & \{d,e\} & \{d\} \\
 & c & \{c,g\} & \{b,c,f,g\} & \{c,g\} \\
 & b & \{b,d,f\} & \{b\} & \{b\} \\
 & a & \{a\} & \{a,d,e,f\} & \{a\} \\
\hline
\end{tabular}
\end{minipage}
\end{table}
\subsection{MRGanter+}
\label{subsec:MRGanter+}
NextClosure calculates closures in lectic ordering to ensure every concept appears only once. This approach allows a single concept to be tested with the closure validation condition during each iteration. This is efficient when the algorithm runs on a single machine. For multi-machine computation, the extra computation and redundancy resulting from keeping  only one concept after each iteration across many machines is costly. We modify NextClosure to reduce the number of iterations and name the corresponding distributed algorithm, MRGanter+.

Rather than using redundancy checking, we keep as many closures as possible in each iteration; All closures are maintained and used to generate the next batch of closures. To this end, we modify  Algorithm~\ref{reducetaskforMRGanter}: the Map algorithm remains the same as in Algorithm~\ref{maptaskforMRGanter}. Algorithm~\ref{reducetaskforMRGanterPlus} describes the ReduceTask for MRGanter+.
The Reduce in MRGanter+ first merges local closures in Line \ref{mrganterplus_merging}, and then recursively examines if they already exist in the set of global formal concepts \mbox{H} (Line \ref{mrganterplus_check}). The set \mbox{H} is used to fast index and search a specified closure, and it is designed as a two-level hash table to reduce its costs. The first level is indexed by the head attribute of the closure, while the second level is indexed by the length of the closure. The new closures are stored in \mbox{G}.
We present a running example based on the dataset in Table~\ref{distributeddataset} for the purpose of comparison. MRGanter+ produces many intents in each iteration. New intents are kept if they are not already in H. Notably, MRGanter+ requires $3$ iterations to mine all concepts.

\section{Twister MapReduce}
\label{sec:twister}
The MapReduce framework adopts a divide-conquer strategy to deal with huge datasets and is applicable to many classes of problems \cite{mr2006ml}.  A large number of computers, collectively referred to as a cluster, are used to run the algorithm in a distributed way.

MapReduce was inspired by the map and reduce functions commonly used in functional programming, for example Lisp. It was introduced by Google \cite{MapReduce_Google} and then implemented by many companies (Google, Yahoo!) and organizations (Twister, Apache). These implementations provide automatic parallelization and distribution, fault-tolerance, I/O scheduling, status and monitoring. The only demand made of the user is the formulation of the problem in terms of map and reduce functions. We use the terminology \emph{mapper} and \emph{reducer} when we refer to the map and reduce function respectively. The map function takes an input pair and produces a set of intermediate key/value pairs. The MapReduce library provides the ability to acquire input pairs from files or databases which are stored in distributed way. Additionally, it can group all intermediate values associated with the same intermediate key \textit{I} and pass them to the same reducer. The reduce function accepts an intermediate key \textit{I} and a set of values associated with \textit{I}. It merges these values to form a possibly smaller set of values.

Twister \cite{twister10} was designed to enhance MapReduce's functionality by efficiently supporting iterative algorithms. Twister uses a public/subscribe messaging infrastructure (we choose NaradaBrokering\footnote{http://www.naradabrokering.org/}) for communication and data transfer, and introduces long running map/reduce tasks which can be re-used in different iterations. These long running tasks, which last for the duration of the entire computation, ensures that Twister avoids reading static data in each execution of MapReduce; a considerable saving. For iterative algorithms, Twister categorizes data as being either static or dynamic. Static data is the distributed data in local machines. Dynamic data is typically the data produced by the previous iteration. Twister's  \emph{configure} phase allows the specification of where the mapper reads the static data. Calculation is performed cyclically based upon the dynamic and static data.

Unlike Twister, Hadoop focuses on single step MapReduce and lacks built-in support for iterative programs. For iterative algorithms, Hadoop MapReduce chains multiple jobs together. The output of a  previous MapReduce task is used as the input for the next MapReduce task\footnote{http://hadooptutorial.wikispaces.com/Iterative+MapReduce+and+Counters}. This approach is suboptimal; it incurs the additional cost of repetitively applying MapReduce --the disadvantage is that new map/reduce tasks are created repetitively for different iterations. This incurs considerable performance overhead costs.

\section{Evaluation}
\label{section:experiments}
We provide evidence of the effectiveness and scalability of our algorithm in this section. Subsection \ref{subsec:env} describes the experimental environment and the dataset characteristics for the datasets used to validate performance in this work. In subsection~\ref{subsec:analysis}, we describe our experimental results.
\subsection{Test Environment and Datasets}
\label{subsec:env}
MRGanter and MRGanter+ are implemented in Java using Twister runtime as the distributed environment. In addition, a distributed version of CloseByOne proposed by Krajca and Vychodil \cite{ClosebyOne1993Kuznetsov} is implemented under the Twister model in order to provide a fair comparison for the algorithms proposed in the present paper. For convenience, we name this algorithm MRCbo. To illustrate the performance improvement of our distributed approach, we also evaluate NextClosure and CloseByOne.

The experiment were run on the Amazon EC2 cloud computing platform. We used High-CPU Medium Instances which had 1.7 GB of memory, 5 EC2 Compute Units (2 virtual cores with 2.5 EC2 Compute Units each), 350 GB of local instance storage, and a 32-bit platform. We selected 3 datasets from UCI KDD machine learning repository, mushroom, anon-web, and census-income for this evaluation\footnote{http://archive.ics.uci.edu/ml/index.html}. These datasets have 8124, 32711, 103950 records and 125, 294, 133 attributes respectively. We used the percentage of 1s to measure the dataset  density (see row 4 in Table~\ref{ucids}). CPU time was used as the metric for comparing the performance of the algorithms. The number of iterations used by each algorithm was also recorded in Fig.~\ref{iterations}.
\begin{table}[t]
\caption{UCI dataset characteristics. These characteristics include the numbers of objects and the number of attributes, and the  density.}
\vspace{-2mm}
\begin{center}
\begin{tabular}{|c|c|c|c|}
\hline Dataset & mushroom & anon-web & census-income \\ 
\hline objects & 8124 & 32711 & 103950 \\ 
\hline attributes & 125 & 294 & 133 \\ 
\hline density & 17.36\% & 1.03\% & 6.7\% \\ 
\hline 
\end{tabular}
\end{center}
\label{ucids}
\end{table}

\subsection{Results and Analysis}
\label{subsec:analysis}
\begin{table}[t]
\centering
\caption{Execution time (in seconds) for each algorithm on the datasets.}
\begin{tabular}{r||r|r|r}
\hline
	Dataset & mushroom & anon-web  & census-income \\
	concepts & 219010 & 129009 & 96531 \\ 
\hline
	NextClosure & 618  & 14671 & 18230  \\ 
	CloseByOne & 2543 & 656 & 7465\\ 
	MRGanter & 20269(5 nodes) & 20110 (3 nodes) & 9654 (11 nodes)  \\ 
	MRCbo & 241 (11 nodes) & 693 (11 nodes) & 803 (11 nodes)  \\ 
	MRGanter+ & 198 (9 nodes) & 496 (9 nodes) & 358 (11 nodes) \\ 
\hline 
\end{tabular}
\label{result}
\end{table}

\begin{figure}[t]
\begin{minipage}[t]{0.5\linewidth}
\centering
\includegraphics[height=.23\textheight]{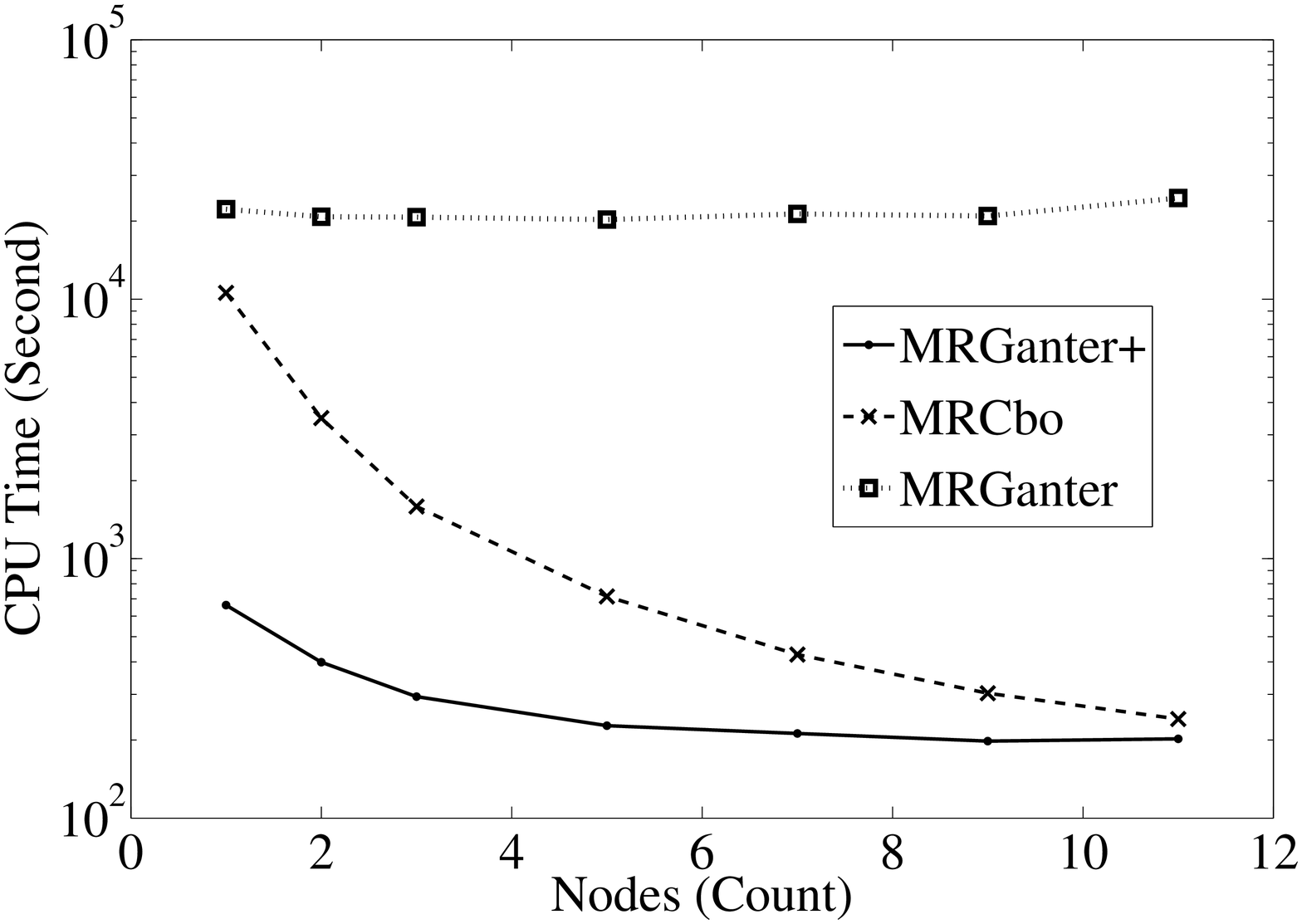}
\caption{Mushroom dataset: comparison of MRGanter+, MRCbo and MRGanter. MRGanter+ outperforms MRCbo and MRGanter when dense data is processed.}
\label{fig:MRmushroom}
\end{minipage}
\hspace{0.15cm}
\begin{minipage}[t]{0.5\linewidth}
\includegraphics[height=.23\textheight]{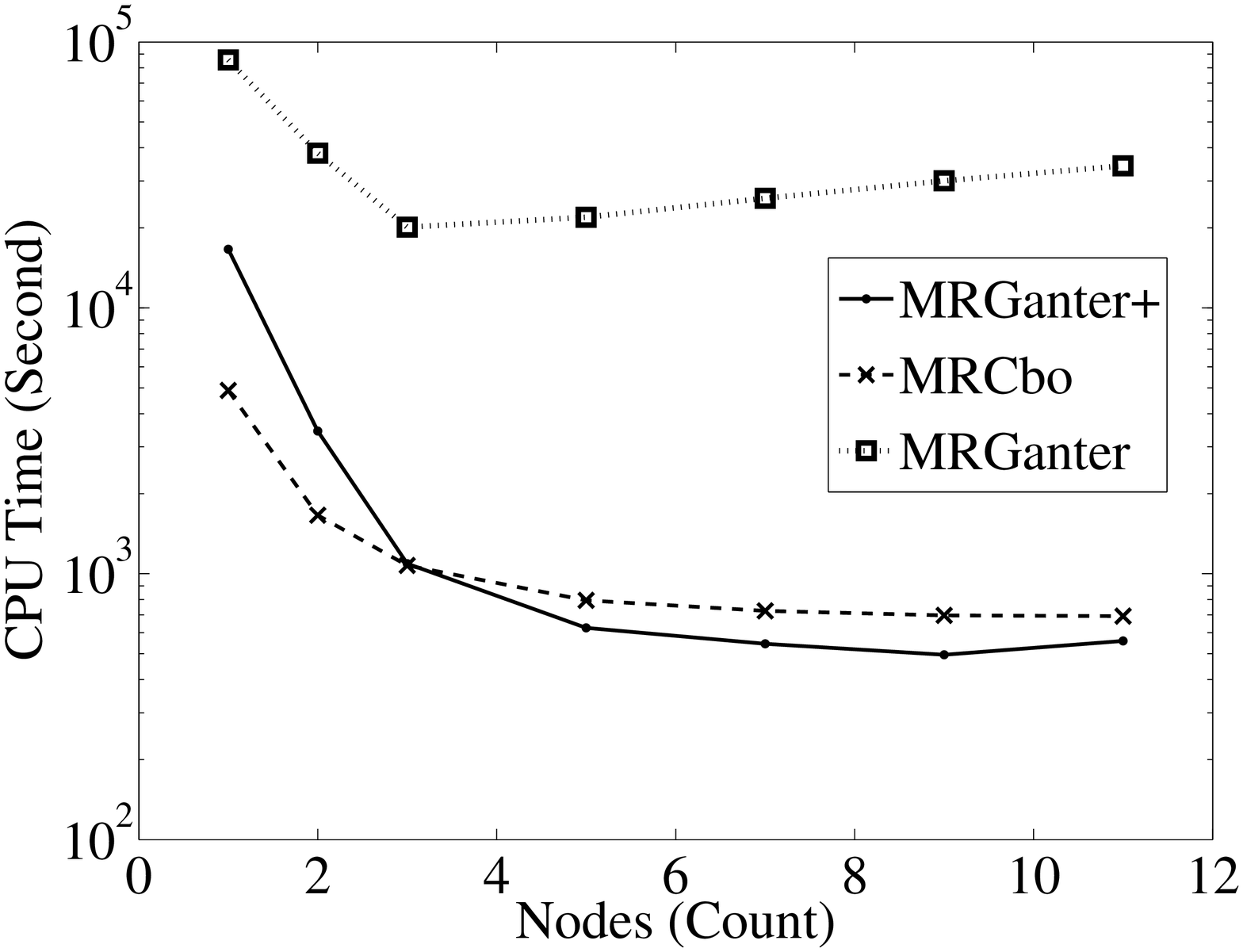}
\caption{Anon-web dataset: comparison of MRGanter+, MRCbo and MRGanter. MRGanter+ is faster when more than 3 nodes are used.}
\label{fig:MRweb}
\end{minipage}
\end{figure}
In Table~\ref{result}, we present the best test results for the centralized algorithms, NextClosure and CloseByOne, and the distributed algorithms, MRGanter, MRCbo and MRGanter+. In short, it is clear that MRGanter+ has the best overall performance for the mushroom, anon-web and census datasets when 9 nodes and 11 nodes are used respectively. In comparison with NextClosure, MRGanter+ saves 68\%, 96.6\% and 98\% in time when processing mushroom, anon-web and census-income dataset respectively. For census-income, MRGanter+ has the best performance. MRGanter+ runs 102 times faster than MRGanter and 1.4 times faster than MRCbo. MRCbo runs much faster than CloseByOne when $11$ nodes are used. It presents a 90.5\% saving in time when dealing with the mushroom dataset compared with CloseByOne, but there is not much of difference when the anon-web dataset is processed. MRGanter takes the longest time to calculate the formal concepts for both the mushroom and anon-web datasets. It is much slower than even the centralized version, NextClosure. The census-income dataset is an exception because MRGanter saves up to half the time with 11 nodes. Among the $\mbox{MR}^{*}$ algorithms and centralized algorithms, MRGanter+ achieved the best performance.

\begin{figure}[t]
\begin{minipage}[t]{0.50\linewidth}
\centering
\includegraphics[height=.23\textheight]{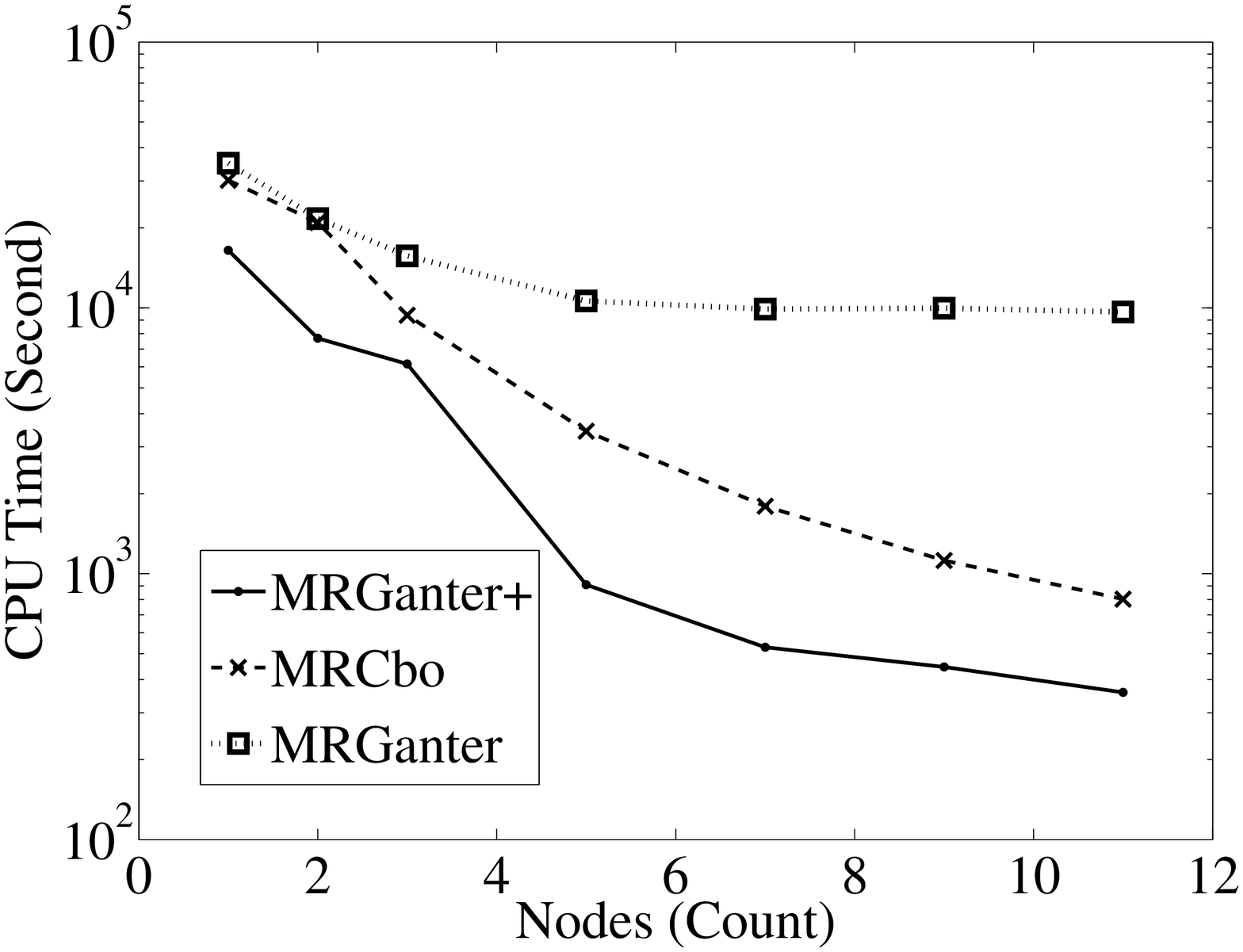}
\caption{Census dataset: comparison of MRGanter+, MRCbo and MRGanter. MRGanter+ is fastest when a large dataset is processed.}
\label{fig:MRcensus}
\end{minipage}
\hspace{0.15cm}
\begin{minipage}[t]{0.48\linewidth}
\vspace{-45mm}
\tabcaption{Number of iterations required for each of the three datasets.}
\label{iterations}
\begin{center}
\begin{tabular}{r|l|l|p{1cm}}
\hline
 Dataset & mushroom
      & anon-web & census-income \\
      \hline
 concepts & 219010
      & 129009 & 96531 \\ 
\hline
 NextClosure & 219010
      & 129009 & 96531 \\ 
 CloseByOne & 14 &
      11 & 11\\ 
 MRGanter & 219010
      & 129009 & 96531 \\ 
 MRCbo & 14 & 11
      & 11 \\ 
 MRGanter+ & 12 &
      11 & 9 \\ 
\hline
\end{tabular}
\end{center}
\end{minipage}
\end{figure}

To go deep into analysis, let us take scalability into account. We tested $\mbox{MR}^{*}$ algorithms on a range of nodes and plotted curves for each of them to show the ability of the algorithms to decrease computation time by utilizing more computers, as indicated in Fig. \ref{fig:MRmushroom}, \ref{fig:MRweb} and \ref{fig:MRcensus} for the different datasets.

In Fig.~\ref{fig:MRmushroom}, MRCbo is slower than MRGanter+ although this curve decreases faster than MRGanter+ when we increase the number of nodes. The execution time of MRGanter+ is fast even on a single node and the execution time keeps decreasing up to the maximum number of nodes, 11. The performance of MRGanter is not beneficially affected by increasing the number of nodes. One explanation for this is the overhead incurred by distributing the computation, in particularly network communication overhead. This is markedly different from MRGanter+, because MRGanter+ produces substantially more intermediate data than MRGanter and MRCbo. Secondly, there is additional computation involved in the distributed algorithms in comparison with the centralized versions of these algorithms. Consider, for instance, the extra operation needed by the merging operation. The best number of nodes, where best refers to performance speed, depends on the characteristics of the dataset.

Fig.~\ref{fig:MRweb} demonstrates that MRGanter+ outperforms MRGanter for the anon-web dataset. One  reason for this performance improvement is that MRGanter+ produces more concepts during each iteration than MRGanter. Fig.~\ref{iterations} indicates that MRGanter+ requires 12, 11 and 9 iterations for each of the datasets, whereas MRGanter requires 219010, 129009 and 96531 iterations to obtain all concepts. These additional iterations incur higher network communication costs. Fig.~\ref{fig:MRcensus} demonstrates that this is also the case for the census dataset. In addition, the curves in Fig.~\ref{fig:MRcensus} are steeper than the curves in Fig.~\ref{fig:MRmushroom} and \ref{fig:MRweb}. These figures give evidence that the performance of the $\mbox{MR}^{*}$ algorithms is related to size and density of the data. Based on these results we posit that $\mbox{MR}^{*}$ algorithms scale well for large and sparse datasets. This evidence suggest that $\mbox{MR}^{*}$ algorithms may be a viable candidate tool for handling large datasets, particularly when it is impractical to use a traditional  centralized technique.

Classical formal concept computing methods usually act on, and have local access to the entire database. Network communication is the primary concern when developing distributed FCA approaches: Frequent requests to remote databases incur significant time and resource costs. Performance improvements of the algorithms proposed in this paper may potentially arise from preprocessing the dataset so that the dataset is partitioned in a more optimal manner. One direction for improving these algorithms lies in making the partitions more even, in terms of density, so that the complexity is distributed more equably. We also intend to extend these methods so that they reduce the size of intermediate data produced in each iteration. We propose to extend this empirical study in a companion paper which examines algorithm performance on larger dataset sizes. 

\section{Conclusion}
\label{sec:conclusion}
In this paper we considered methods for extending the NextClosure FCA algorithm. A formal description of dealing with distributed datasets for the NextClosure FCA was discussed. Two new distributed FCA algorithms, MRGanter and MRGanter+, were proposed based on this discussion. Various implementation aspects of these approaches were discussed based on empirical evaluation of the algorithms. These experiments demonstrated the advantages of our approach and the scalability in particular of MRGanter+. By comparing MRGanter+ with MRCbo and MRGanter, we found that the number of iterations significantly impacted the performance of distributed FCA, a promising result. In future work we hope to capitalize on this by improving the  $\mbox{MR}^{*}$ methodology by reducing the number of iterations of these approaches and to further reduce computation time. 

\section{Acknowledgement}
This paper is the author's self-archiving version of the original publication. It is only allowed to use the content for non-commercial and internal educational purposes. \href{http://www.springerlink.com/content/02p8282703rx0m78/}{The original publication is available at www.springerlink.com}

\bibliography{Bib_BX}{}
\bibliographystyle{unsrt}
\end{document}